\documentclass{cupconf}
\usepackage{epsfig}

  \checkfont{eurm10}
  \iffontfound
    \IfFileExists{upmath.sty}
      {\typeout{^^JFound AMS Euler Roman fonts on the system,
                   using the 'upmath' package.^^J}%
       \usepackage{upmath}}
      {\typeout{^^JFound AMS Euler Roman fonts on the system, but you
                   dont seem to have the}%
       \typeout{'upmath' package installed. cupconf.cls can take advantage
                 of these fonts,^^Jif you use 'upmath' package.^^J}%
      }
  \else
  \fi

  \checkfont{msam10}
  \iffontfound
    \IfFileExists{amssymb.sty}
      {\typeout{^^JFound AMS Symbol fonts on the system, using the
                'amssymb' package.^^J}%
       \usepackage{amssymb}%
         \let\leq=\leqslant
         
      }{}
  \fi

  \IfFileExists{amsbsy.sty}
    {\typeout{^^JFound the 'amsbsy' package on the system, using it.^^J}%
     \usepackage{amsbsy}}
    {}

\newsavebox{\astrutbox}
\sbox{\astrutbox}{\rule[-5pt]{0pt}{20pt}}

\title[Making black holes visible]{Making black holes visible: accretion,
radiation, and jets}

\author[J.H. Krolik]
{J\ls U\ls L\ls I\ls A\ls N\ns H.\ns K\ls R\ls O\ls L\ls I\ls K$^1$}{}

\affiliation{$^1$Department of Physics and Astronomy, Johns Hopkins University,
Baltimore, MD 21218, USA}

\date{August 9, 2007 and in revised form ??}
\begin{document}

\maketitle

\begin{abstract}

     With the fundamental stress mechanism of accretion disks identified---correlated
MHD turbulence driven by the magneto-rotational instability---it has become possible
to make numerical simulations of accretion disk dynamics based on well-understood
physics.  A sampling of results from both Newtonian 3-d shearing box and general
relativistic global disk MHD simulations is reported.   Among other things, these
simulations have shown that: contrary to long-held assumptions, stress is continuous
through the marginally stable and plunging regions around black holes, so that
rotating black holes can give substantial amounts of angular momentum electromagnetically
to surrounding matter; the upper layers of accretion disks are primarily
supported by magnetic pressure, potentially leading to interesting departures
from local black-body emitted spectra; and initially local magnetic fields in
accretion flows can, in some cases, spontaneously generate large-scale fields
that connect rotating black holes to infinity and mediate strong relativistic jets.

\end{abstract}

\firstsection 
\section{Prolog: the classical view of accretion disks}

     It has been understood for decades that accretion through disks can
be an extremely powerful source of energy for the generation of both photons
and material outflows.   When the central object is a black hole, the gravitational
potential at the center of the disk is relativistically deep, so that the
amount of energy that might be released per unit rest-mass accreted can
be a substantial fraction of unity.  If the central black hole spins, an
additional store of tappable energy resides in its rotation.

    At the same time, however, the physical processes by which matter inflow
is transmuted into observable outputs has long remained extremely murky.  For matter
to move inward, it must somehow lose its orbital angular momentum.  That there
might be some sort of inter-ring friction seems plausible, given the orbital
shear, but the way this frication is generally envisaged is in terms of
an imaginary ``viscosity coefficient" famously parameterized by \cite{ss73}
as $\alpha c_s h$, for local sound speed $c_s$ and vertical scale-height $h$.
This {\it ansatz} is based entirely on dimensional analysis: the (unknown)
local stress is set equal to a dimensionless number $\alpha$ times the
local pressure solely because the local pressure has the same units as stress.
Although there is no particular reason why the stress, measured in local
pressure units, should always have the same value, $\alpha$ is very frequently
assumed to be a constant at all places and at all times.  Moreover, despite
the fact that ordinary molecular viscosity fails miserably to explain the
friction, it is often assumed that the stress is some sort of intrinsically
dissipative kinetic process whose operation is at least analogous to that
of conventional viscosity.

     Inter-ring torques do work, moving energy outward.  Simultaneously, matter
moves inward, carrying its orbital energy.  In a steady-state disk, where the
mass inflow rate is the same at all radii, these two energy fluxes do not cancel.
Instead, there is a net amount of energy that must be deposited in each
ring.  If one thinks of the torques as due to some sort of kinetic process
like viscosity, it is natural to suppose that this energy imbalance
is deposited as heat.
The radial profile of this heating in a steady-state disk can be easily
written down when the disk is in steady state, provided one is able to guess
a boundary condition at the inner edge of the disk (we will return to this issue
in greater detail
later).  Integrating over the radial heating profile then immediately yields
the total amount of energy per accreted rest-mass that could, in principle,
be used for radiation, i.e., the radiative efficiency.  Unfortunately, its
value depends strongly on the guessed boundary condition.

   When the matter density is large enough (as it often should be), atomic collisional
processes can efficiently transform the heat into photons, which can then
escape after diffusing vertically through the disk.  In conditions of high
density and large optical depth, the emergent spectrum should be nearly
thermal, an argument that has led many to assume that the spectrum is locally
Planckian.  Note the use of ``it is natural to suppose" and ``can" and ``should be"
and generic terminology such as ``collisional
processes" here; the specific mechanisms for all these steps are as little
understood as the actual torque is when analyzed through the $\alpha$-model.

     How exactly to make use of the black hole's rotational energy has been
in a similarly unsatisfactory state.  Although \cite{bz77} pointed out thirty
years ago that magnetic fields can in principle efficiently convey black hole
rotational energy from deep within
the black hole's ergosphere all the way to infinity, knowledge about this
mechanism's details has been almost as scanty as for accretion.  The particular
solution found by Blandford and Znajek and extended by \cite{ph83} assumes that
the magnetic field is essentially force-free, with (almost) negligible matter inertia
everywhere (including in the plane of the accretion disk) and is valid only
for small spin parameter $a/M$.  In addition, nothing in that theory specifies
the strength of the magnetic field, or explains how it came to extend to infinity.

\section{Genuine disk physics}

     I would not have painted such a gloomy picture if I did not intend
quickly to adopt a
very different attitude and report on some very significant recent progress.  As
a result of this progress, many of the mysteries bemoaned in the preceding
section can now be substantially solved by the application of well-understood
physical processes.   In some cases, the main stumbling block is not lack of
physics knowledge but lack of computing power.  With this recently-gained
understanding, it has become possible to outline a program by which the entire
process, from mechanics of mass inflow to photon generation to jet launching,
might reasonably be followed as a sequence of connected events.

     The story behind these advances begins fifteen years ago when Steven Balbus
and John Hawley pointed out that weak magnetic
fields destabilize an orbiting disk and lead to the rapid growth of MHD turbulence
(\cite[Balbus \& Hawley 1991]{bh91a};\cite[Hawley \& Balbus 1991]{bh91b}).  Orbital
shear is what drives this ``magneto-rotational instability";
orbital shear also enforces a correlation in the resulting MHD turbulence such that the
magnetic stress $-B_r B_\phi/4\pi$ is always on average positive.  That is, the shear
itself ensures that the magnetic forces transmit angular momentum outward.

     It is important to note that magnetic stress, unlike viscosity, is {\it not}
intrinsically dissipative.  The local rate of heat generation is {\it not} always
proportional to the local stress.  On the other hand, it is intimately connected
to dissipation.  Nonlinear mode-mode interactions can convey turbulent energy
from the relatively large lengthscales on which the turbulence is stirred to
much shorter lengthscales on which a variety of genuinely dissipative processes
can act efficiently.  When integrated over a large enough volume (vertically-integrated
through the disk and wide enough to make averaging well-defined), the heating rate
must agree with the one predicted by the time-steady disk picture described above,
but there is no requirement for it to match up with the stress rate locally,
either in time or space.

     The radiative output can then be described by taking a position- and time-dependent
heating rate from turbulence calculations and solving the radiation transfer
equation using physical opacities.  If the vertical structure of the disk is
known well-enough, non-LTE effects in the disk atmosphere can be incorporated,
and departures from locally Planckian spectra can be predicted.

     To find the global radial profile of heating and radiation, it is still
necessary to understand that inner boundary condition.  Thirty years ago, heuristic
arguments based on hydrodynamic intuition pointed toward a boundary condition
requiring the stress to be zero at and inside the marginally stable orbit, but
even then it was recognized that if magnetic fields were important, a different
choice might be necessary (\cite[Page \& Thorne 1974]{pt74}).  Now that we
know magnetic fields are essential to the entire accretion process, what Page
and Thorne saw as a back-of-the-mind worry is now front-and-center.  Fortunately,
as we shall shortly see, it is now possible to {\it calculate}, rather than guess,
what happens to the magnetic stress in the marginally stable region.

      If we can compute the magnetic field in the marginally stable region,
then we can also compute the magnetic field even closer to the black hole's
event horizon.  Part of its nature is determined by dynamics within the
accretion flow itself; part may be constrained by boundary conditions at
infinity, where it is possible that some fieldlines are anchored.

\section{Numerical simulations}

     The only fly in the ointment is that analytic methods for calculating
the properties of any turbulent system are extremely limited in their power.
Numerical simulation is really the only tool we have for examining fully-developed
turbulence, and it has its own limitations.   Nonetheless, some fifteen years
after the first attempt to study numerically the properties of MHD turbulence
in accretion disks, a great deal has been learned.

     The simulations that have been done to date can all be divided into two
classes: shearing boxes and global disks.   To make a shearing box, imagine cutting
out from a complete disk a narrow radial annulus of limited azimuthal extent.
When its azimuthal length is small compared to a radian, it can be well
approximated as straight along the tangential direction.  Rather than describing
the orbital motion by a rotational frequency $\Omega(r) = \Omega_0 (r/r_0)^{-3/2}$
for radius $r$ and annular central radius $r_0$, we can instead approximate
it by writing the orbital speed as $v_y = r_0 \Omega_0\left[1 - (1/2)(r/r_0 - 1)\right]$,
for $r/r_0 - 1 \ll 1$.   The equations of motion can then
be written in the rotating frame with appropriate centrifugal and Coriolis
terms.  At a similar level of approximation, the vertical gravity is
$g_z = z \Omega_0^2$.

    Shearing boxes are best for wide dynamic range studies of the
turbulent cascade, well-resolved exploration of internal vertical structure
within the disk, and tracing disk thermodynamics.   The advantage for
the latter subject is that the relatively large dynamic range in lengthscale
for turbulent dynamics allows one to localize comparatively well where
dissipation occurs and then follow the diffusion of radiation away from
its source regions.

      In a global disk simulation, one places a large amount of mass on
the grid in an initial state of hydrodynamic equilibrium.  To avoid
noise propagation across the boundary of the problem area, the outer
boundary is placed well outside the outer edge of the initial
mass distribution.  When angular momentum begins to flow outward
through the disk as a result of the MHD turbulent torques, matter
from the inner part moves inward while a small amount of mass on
the outside moves outward, soaking up the angular momentum that
has been carried outward to it.  Thus, these global disks are
truly ``accretion disks" only in their inner portions.  For this
reason, it is necessary to locate their initial centers far enough
outside the innermost stable circular orbit (the ISCO, i.e., the
radius of marginal stability) that there can be a reasonable
radial dynamic range for the accretion flow proper.

     In the current state-of-the-art, shearing box simulations employ
3-d Newtonian dynamics in the MHD approximation including radiation
forces.  By means of integrating both an internal energy and a total
energy equation, they can track the numerical dissipation rate as
a function of time in each cell of the simulation; this numerical
dissipation rate is assumed to mimic the physical dissipation and
is used to increase the heat content of the cells where and when
it occurs.  Radiation transfer is computed in the approximation
of flux-limited diffusion, using thermally-averaged opacities
(\cite[Hirose et~al. 2006]{hks06}).

     Global disk simulations are best for following the inflow
dynamics, the radial profile of magnetic stress, the surface density
profile that the stress produces, and non-local magnetic
field effects.  They also permit identification of typical
global structures (the main disk body, disk ``coronae", etc.)
and the study of jets.

     The most physically complete global disk simulations now
available use 3-d fully general relativistic dynamics in the
MHD approximation, but they take no account of radiation.  In
one version (\cite[Gammie et~al. 2003]{gmt03}), the total energy
equation is integrated, so energy is rigorously conserved; in another
(\cite[De~Villiers \& Hawley 2003]{deVH03}), an internal
energy equation is solved, permitting numerical energy losses.
The advantage of the former method is that numerical dissipation
does not lead to energy loss; its disadvantage is that physical
radiation losses don't occur either.   The advantages and
disadvantages of the latter method are more or less reversed,
if one is willing to accept numerical energy losses as
approximating genuine radiative losses.

     In all cases, both shearing box and global simulations,
the magnetic field is nearly always assumed to have zero net
flux.  This choice is made largely because it's the simplest
and involves the smallest number of arbitrary choices: the
initial field on the boundary of the simulation is always zero.
On the other hand, it is possible that there can be large-scale
fields running through real accretion flows, and they may have
substantial effects on the character of those flows; this
question remains to be investigated.

\section{Selected Results}

    The body of this talk will be devoted to a brief summary of
some of the principal achievements of these simulations.  Consistent
with my title, I will focus on three topics: what we have learned
from the global simulations about the radial profile of stress
(and possibly of dissipation); what shearing box simulations have
shown us about the vertical structure of disks, and how that
can influence the character of the emitted spectrum; and how
magnetically-driven accretion can (or maybe not) launch relativistic
jets.

\subsection{Radial stress profiles}

    The De~Villiers-Hawley simulation code is designed to do an
excellent job of conserving angular momentum and propagating
magnetic fields reliably, but is less good
at conserving energy.  Consequently, we believe its description
of the electromagnetic stress and its relation to mass inflow should
be reliable, but it is much more difficult to use the data from
these simulations to predict dissipation and the radiation
that may follow from it.  For the time being, then, we can
discuss the stress with some confidence, while using it as an
indirect and approximate indicator of dissipation.

\begin{figure}
  \centerline{\psfig{file=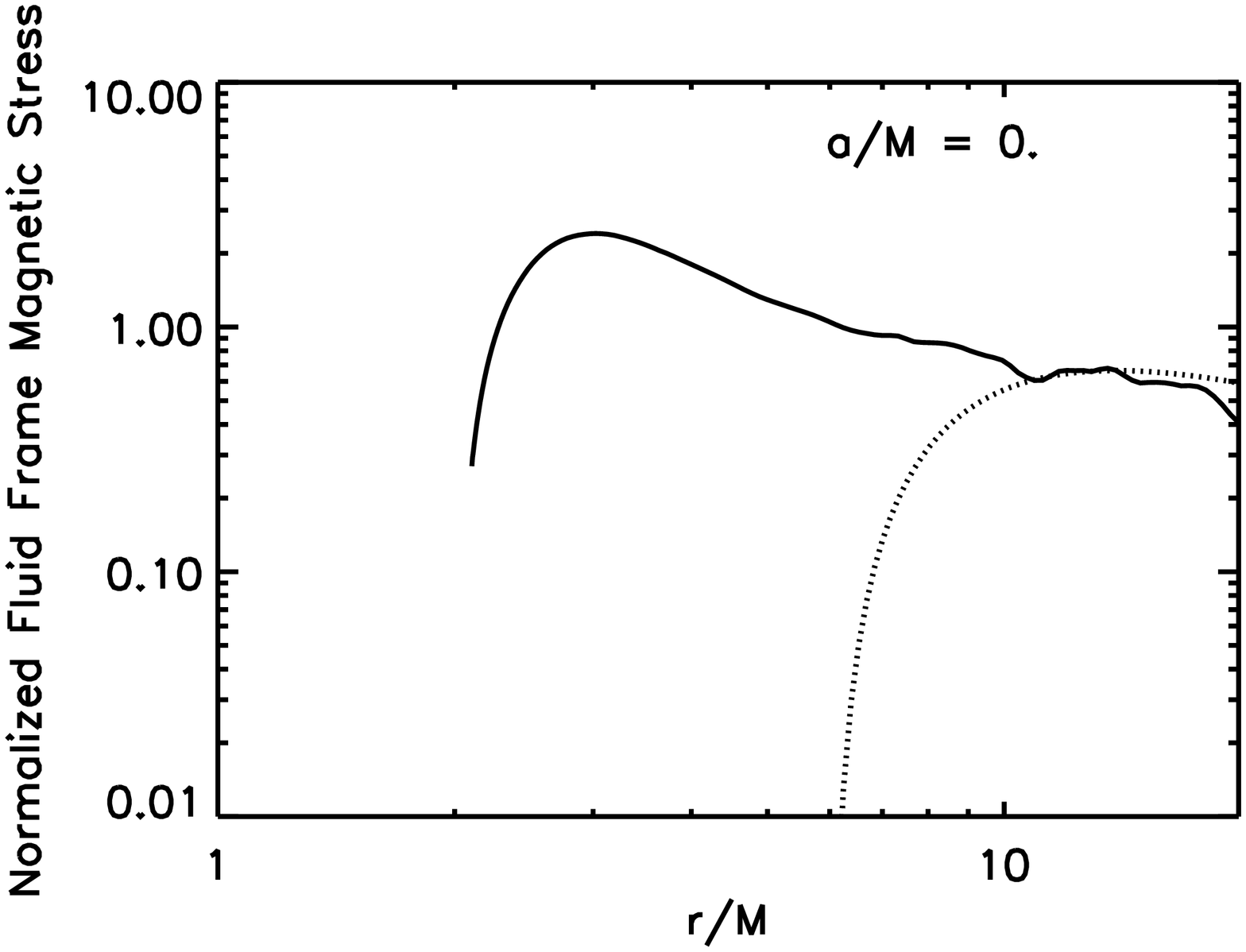,angle=90,width=4.0in}}

  \centerline{\psfig{file=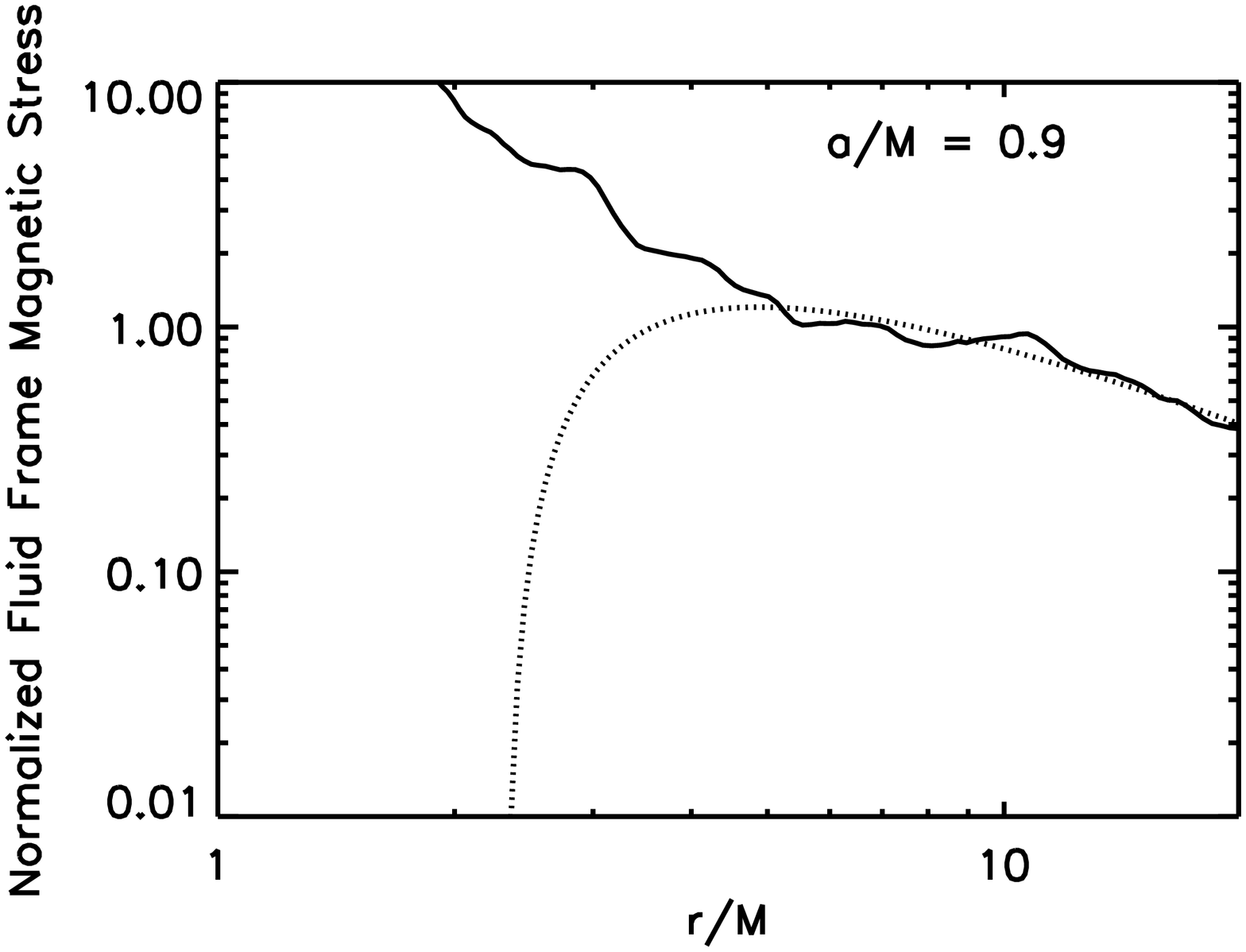,angle=90,width=4.0in}}

  \caption{Fluid-frame stress, integrated over spherical
  shells.  Upper panel shows a simulation with a non-rotating black hole,
  lower panel one with a black hole having a spin parameter $a/M = 0.9$.
  Solid curves are the electromagnetic stress as found in the simulations at
  a particular time; dotted curves are the prediction of the Novikov-Thorne
  model for an accretion rate matching the time-average of that simulation.
  \label{fig:stress}}
\end{figure}

     Figure~\ref{fig:stress} shows the instantaneous shell-integrated
electromagnetic stress (i.e., $-b^r b_\phi + ||b||^2 u^r u_\phi$, for magnetic
four-vector $b_\mu$ and four-velocity $u_\mu$) evaluated in the local fluid-frame
at late-times in two simulations, one with a non-rotating black hole, the other
with a black hole having spin parameter $a/M = 0.9$.   For comparison, the
figure also shows the stress predicted by the Novikov-Thorne model (i.e.,
a time-steady disk with a zero-stress inner boundary condition) when the
accretion rate is the same as the time-averaged value in the corresponding
simulation.   As can readily be seen, the zero-stress boundary condition
fails drastically to describe what actually happens.   In both cases,
the electromagnetic stress continues quite smoothly inward through the
marginally stable region.  In the Schwarzschild case, the stress rises
slowly inward until just outside the event horizon, and then plummets
as the event horizon is approached.  In the Kerr case, the stress rises
sharply upward and does not diminish even very near the horizon.

     It is easy to understand qualitatively both stress profiles.  Because
there is no reason
for the magnetic field to disappear suddenly at the ISCO, while orbital
shear continues to stretch any radial components in the azimuthal direction,
there is no physical mechanism to eliminate magnetic stress there
(\cite[Krolik 1999]{k99}, \cite[Gammie 1999]{g99}); rather, the
stress continues and, if anything, strengthens.  The contrast in behavior
between the spinning and non-spinning cases can just as easily be understood
when one thinks of stress as momentum flux.  The electromagnetic stress
is nothing else than an outward flux of angular momentum, carried in the
electromagnetic field.   A non-rotating black hole has no angular momentum
to give up, so it cannot act as a source for outgoing electromagnetic
angular momentum flux; on the other hand, the angular momentum of a rotating
black hole can be tapped, and we see this process in action here.  Those
concerned by an outflow of anything from an event horizon should have their
qualms removed by the recognition that there is nothing to prevent a
rotating black hole from swallowing negative angular momentum, i.e., angular
momentum corresponding to rotation in the opposite sense.  This is, of
course, completely equivalent to releasing positive angular momentum.
Indeed, deep in the ergosphere, the electromagnetic energy-at-infinity
is frequently negative (\cite[Krolik et~al. 2005]{khh05}).

     As previously discussed, although the work done by magnetic stress does
not necessarily correspond to any particular local rate of heating, there is
a relationship on a more globally-averaged level.  Thus, the curves of integrated
fluid-frame stress shown in Figure~\ref{fig:stress} hint
that dissipation may also continue smoothly across the marginally stable
orbit, also contrary to the guessed boundary condition of the Novikov-Thorne
model.

     A further suggestion that this is so comes from a different argument.
Many of the specific physical mechanisms of dissipation in this context are
associated with regions of high current density.  If, for example, there is
a (small: we make the MHD approximation, after all) uniform resistivity $\eta$ in
the fluid, the local heating rate is $\eta ||J||^2$, where $J^\mu$ is the
electric current four-vector and $||J||$ is its scalar magnitude.  In fact,
the dissipation may be even more closely associated with $||J||^2$ than
this simple guess would suggest because there are a number of plasma
instabilities that create anomalous resistivity precisely where the
current is strong.  Maps of the current density show that it is strongly
concentrated toward the center of the accretion flow, rising rapidly
into the plunging region inside the ISCO (\cite{hkdh04}).   To the degree
that current density indicates candidate regions for rapid magnetic energy
losses, this signal, too, suggests that there may be a great deal of
dissipation in and within the marginally stable region.

     The continuation of stress through the plunging region can also
be looked at from a different point of view: in the language of the
Shakura-Sunyeav $\alpha$ model.   Their argument from dimensional analysis
was that the time-averaged vertically-integrated stress should be comparable
to the time-averaged vertically-integrated pressure.   Our data confirm
that this is so, provided one interprets ``comparable" loosely.  In the
disk body, that is, at radii well outside the ISCO and well inside the
initial pressure maximum (beyond which there is no accretion), the time-average
ratio at a single radius of vertically-integrated stress to vertically-integrated
pressure is generally in the range $\sim 0.01$--0.1.   However,
the instantaneous value of this ratio can easily change
by factors of several over an orbital period.

     Moreover, if one tracks this ratio from somewhat outside the ISCO
to well inside, there is a consistent trend for the ratio between stress and
pressure to increase.  As shown in Figure~\ref{fig:stress}, the stress
generally increases inward in this region; because the radial speed of
the accretion flow also increases inward, the vertically-integrated
density and pressure of the
matter tend to decrease.  The result is that the ratio of stress to
pressure increases sharply, often rising by factors of 10--100 from
the disk body to deep inside the plunging region.  Thus, a description
of inflow dynamics near the ISCO in terms of a constant $\alpha$
parameter is strongly in conflict with the results of these simulations.
Claims (as are often made) based on assuming a constant value of $\alpha$
within this region are therefore on very shaky ground.

\subsection{Internal vertical structure}

     Having thus emphasized the consequences of magnetically-driven
accretion in the inner part of the accretion flow, it is now time to turn to
its implications for the internal structure of accretion disks at larger radii.
The best tool for studying this problem is shearing box simulations that
both accurately conserve energy and follow radiation transfer (as described
in \cite[Hirose et~al. 2006]{hks06}).  In this review, we will briefly
discuss two of the pricipal results of these simulations: their implications
for thermal stability in disks and the fact that disk upper layers are
generically supported primarily by magnetic fields.

      In their classic paper on the $\alpha$-model, Shakura and Sunyaev
(1973) also predicted that the inner regions of all disks surrounding
black holes in which the accretion rate is more than a small fraction
of the Eddington rate should be dominated by radiation pressure.  Three
years later (\cite[Shakura \& Sunyaev 1976]{ss76}), the same two authors
demonstrated that, within the approximation scheme of the vertically-integrated
$\alpha$-model, radiation pressure dominance leads directly to thermal
instability.   If so, the standard equilibrium solution for the region from
which most of the light is generated in the brightest accreting black hole
systems is unstable.  To this day, no satisfactory resolution to the
question of, ``What actually happens in these circumstances?" has emerged.

\begin{figure}
  \centerline{\psfig{file=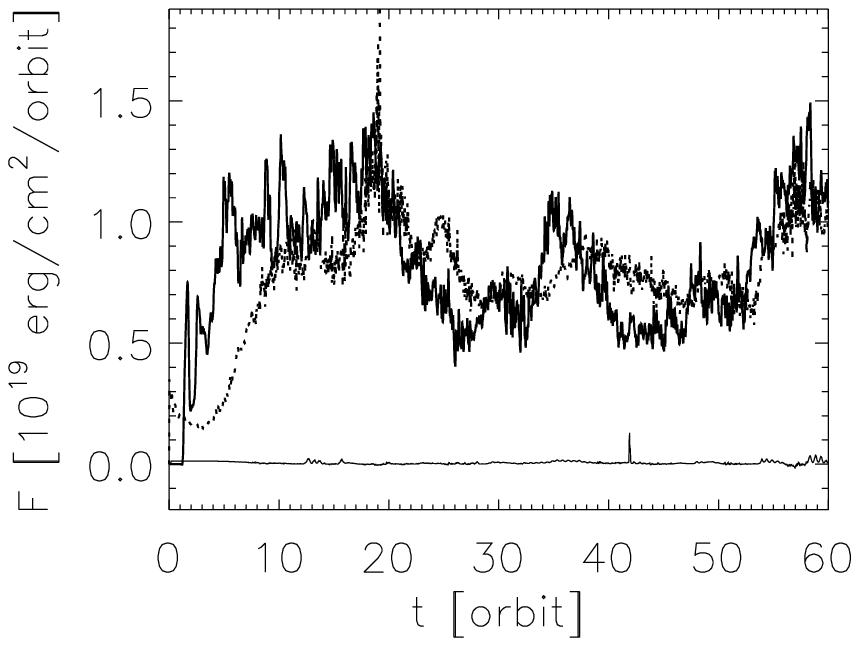,width=2.8in} \quad
              \psfig{file=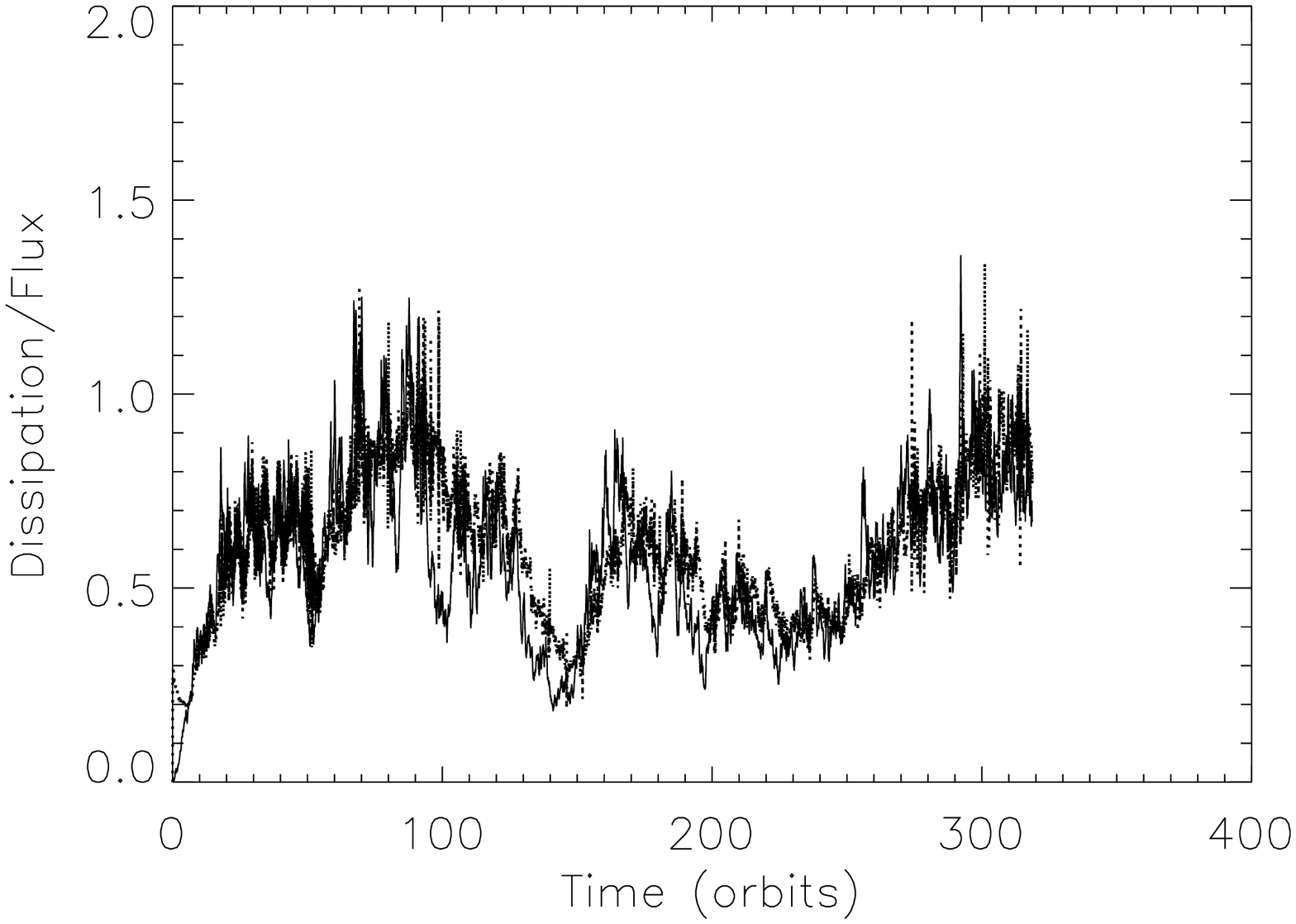,width=2.2in, angle=90}}
  \caption{Heating and radiative output from two shearing box simulations.  Left panel
   shows a case in which $p_r/p_g \simeq 0.2$ (taken from \cite[Hirose et~al.
   2006]{hks06}); right panel a case in which $0.5 \leq p_r/p_g \leq 2$, depending
   on the time within the simulation \cite[Krolik et~al. 2007]{khb07}).
   In both figures, the solid curve shows the volume-integrated dissipation rate,
   while the dashed curve shows the radiative output.  After initial transients,
   heating and radiative flux are nearly identical.
   \label{fig:radoutput}}
\end{figure}

      One of the principal motivations for our program of simulating
shearing boxes with radiation generation and transport is to answer this
question.  At this stage, there is progress to report, but not yet any
firm answer to the big question.   When the gas pressure is dominant,
it is clear that a truly stable steady-state can be found.
Figure~\ref{fig:radoutput}a illustrates this point by showing the
``light-curve" of a shearing box disk segment in which the radiation pressure
$p_r$ is only about $20\%$ of the gas pressure $p_g$.  The fluctuations in
radiative output are only at the tens of percent level.

     On the other hand, increasing radiation pressure does tend to drive
fluctuations.  When the radiation and gas pressures are comparable
(Fig.~\ref{fig:radoutput}b), the output flux varies over a range of
a factor of 3--4.  Intriguingly, although $p_r/p_g$ at its greatest
is above the threshold for instability suggested by Shakura and Sunyaev,
and stays there for as long as 5 cooling times,
the disk exhibits large limit-cycle oscillations, but no unstable runaway.
We are actively investigating whether the instability remains under control
at still higher values of $p_r/p_g$ (Hirose et~al., in preparation).

\begin{figure}
  \centerline{\psfig{file=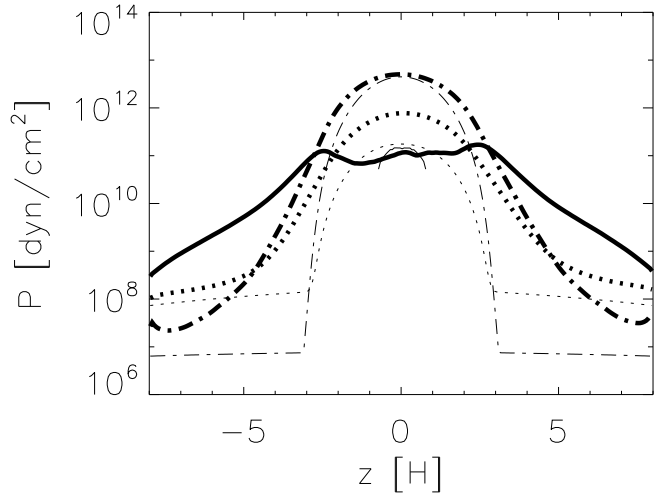,width=4.0in}}
  \caption{Vertical profiles of several kinds of pressure in a simulation
whose volume- and time-averaged ratio of radiation pressure to gas pressure was
$\simeq 0.2$ (\cite[Hirose et~al. 2006]{hks06}).   The thick curves are
time-averaged over fifty orbital periods starting from the end of initial transients;
the thin curves are the initial conditions.   Magnetic pressure is shown by
the solid curves, radiation by the dotted curves, and gas pressure by the
dash-dot curves.
\label{fig:vertstruct}}
\end{figure}

      A consistent result of all vertically-stratified shearing box studies
is that their upper layers are magnetically-dominated (\cite[Miller \&
Stone 2000]{ms00}; and as shown in Fig.~\ref{fig:vertstruct}, taken from
\cite[Hirose et~al. 2006]{hks06}).  Although the data shown here are from
a particular gas-dominated simulation, more recent work studying shearing boxes
with radiation pressure comparable to gas pressure (\cite[Krolik et~al. 2007]{khb07})
and radiation pressure considerably greater than gas pressure (Hirose et~al.,
in preparation) shows very much the same pattern: independent of whether
gas or radiation pressure dominates near the midplane, by a few scale-heights
from the center, magnetic pressure is larger than either one.

    A somewhat surprising corollary of magnetic dominance in the upper
layers is that ``coronal" heating is rather limited.   Strong hard X-ray
emission is so commonly seen from accreting black holes, no matter whether
the central black hole has a mass $\sim 1 M_{\odot}$ or $\sim 10^9 M_{\odot}$,
that somewhere in the system there must be a region of intense heating
with only small matter density and optical depth.  Otherwise, there would
be no way to heat electrons to the $\sim 100$~keV temperatures required to
produce the X-rays.   This region is generally called the accretion disk
``corona", and it is often thought of in conceptual terms derived from experience
with the Solar corona: it is imagined that somehow magnetic field loops
emerge buoyantly from the nearby disk, twist and cross, and release energy
at reconnection points.

   Unfortunately for this popular scenario, these
simulations, the first to treat the dynamics of the upper layers of disks
in a consistent fashion, show no sign of anything resembling it.  The very
fact that magnetic fields dominate the energy density of the upper strata
of disks means that these regions are comparatively quiet, and the strong
magnetic fields themselves prevent any twisting or field line crossing
that might lead to reconnection.  Rather, orbital shear stretches radial
field components into azimuthal field with great regularity and smoothness
(\cite[Hirose et~al. 2004]{hkdh04}).  At the same time, the nonlinearly-saturated
Parker instability (i.e., magnetic buoyancy counter-balanced by magnetic tension)
supports smooth ``plateaus" of field interrupted by occasional narrow cusps
(\cite[Blaes et~al. 2007]{bhk07}).   The smoothness and steadiness of
the magnetic field in this kind of corona tends to suppress dissipation.

\begin{figure}
  \centerline{\psfig{file=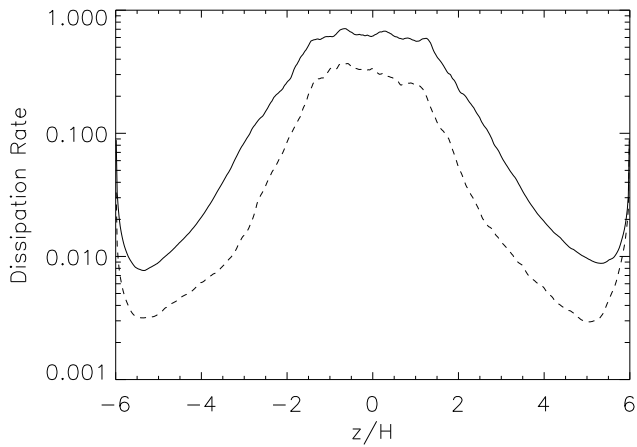,width=2.6in, angle=90}}
  \caption{Time-averaged vertical profiles of the dissipation rate in
  a shearing box simulation.  In this simulation, the mean gas and
  radiation pressures were comparable, but there were order unity
  fluctuations.  The solid curve pertains to those times when the
  radiation presssure was particularly high, the dashed curve to those
  times in which it was particularly low.
  \label{fig:diss_struct}}
\end{figure}

    Indeed, the dissipation in shearing box segments of accretion disks
is typically confined to the central regions of the disk.  Figure~\ref{fig:diss_struct}
(\cite[Krolik et~al. 2007]{khb07}) illustrates this fact in a shearing box
whose radiation and gas pressures were, on a time-averaged basis comparable,
but in which the ratio of radiation to gas pressure fluctuated over the
range 0.5--2.   Whether the energy content of the disk was high (and the
radiation pressure dominated the gas pressure) or low (and the ratio went
the other way), the dissipation was still confined to the inner few scale-heights
of the disk.

\begin{figure}
  \centerline{\psfig{file=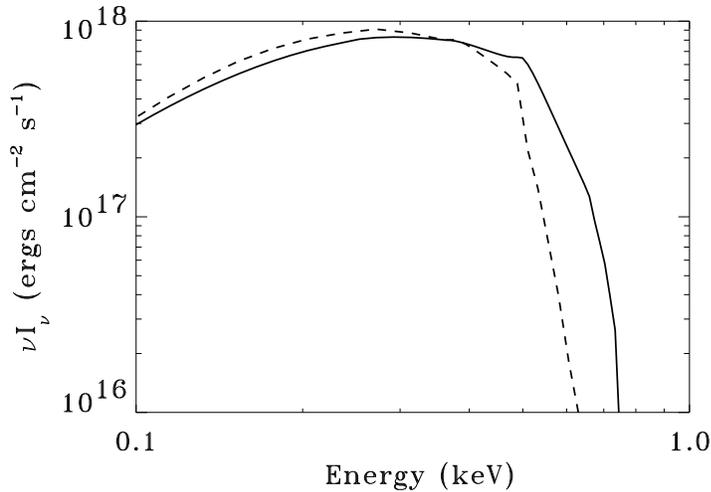,width=4.0in}}
  \caption{Output spectrum from a gas-dominated shearing box at $55^{\circ}$
  from the local vertical direction.  The solid curve shows the spectrum
  predicted when the atmosphere is magnetically-supported, the dashed curve
  shows the predicted spectrum when magnetic support is neglected
  (\cite[Blaes et~al. 2006]{bdhks06}).
  \label{fig:no-bb}}
\end{figure}

     Although magnetic dominance in the upper layers of disks likely does
not lead to strong coronal heating, it does have other potentially important
observational consequences.  Chief among them is the fact that when
magnetic pressure gradients replace gas pressure gradients as the matter's
principal support against gravity, the density of the gas must fall.
Because the photosphere of the disk is located where magnetic support is
so important, we immediately infer that previously-estimated photospheric
densities were too high, and that LTE may not be enforced as thoroughly
as previously thought (\cite[Blaes et~al. 2006]{bdhks06}).  The locally-emitted
spectrum may therefore have larger deviations from Planckian, and if these
features are sufficiently strong, may be visible in the disk-integrated spectrum.
An example is shown in Figure~\ref{fig:no-bb}; when magnetic support
is properly included in the atmosphere structure, a prominent CVI edge appears.
Still further departures from conventional spectral predictions may arise
from the fact that at any given moment, the atmosphere can depart
substantially from the usual picture of plane-parallel symmetry.

     The surprising quietness of the magnetically-dominated regions of
shearing box atmospheres motivates a search for other places to supply
the intense heating required to explain the observed hard X-ray emission.
Better places to look might include the plunging region, which is
both strongly magnetized and highly dynamical, and the region just above
it from which jets are launched.

\subsection{Jet launching}

    For many years, two models have dominated thinking about the launching
of jets: the Blandford-Znajek mechanism (\cite[Blandford \& Znajek 1977]{bz77})
and the Blandford-Payne scheme (\cite[Blandford \& Payne 1983]{bp83}).  The
two models share two central elements: large-scale poloidal magnetic fields
that extend from infinity and pierce the midplane of the accretion flow;
and dynamically-enforced rotation.  They are distinguished by whether the
rotation is enforced by space-time frame-dragging (Blandford-Znajek) or
the orbiting matter of the accretion disk (Blandford-Payne).  Because the
latter depends in an essential way on accretion, whereas the other (at
least in principle) does not, they are also distinguished by whether the
inertia of matter is significant.  Lastly, they differ in the source of
power for the jets: orbital energy of accreting matter for Blandford-Payne,
the rotational kinetic energy of the black hole itself for Blandford-Znajek.

    Both models' dependence on externally-imposed large-scale fields is
problematic because the most natural way to bring magnetic fields into
either the inner parts of accretion disks or all the way to the black hole
event horizon is by advection along with accreted fluid.   However,
we have no way of knowing whether such large-scale connections might survive
the many orders of magnitude in compression suffered by the matter;
reconnection might destroy large-scale connections far from the black
hole.   On the other hand, if even a small fraction of the flux survives,
over time it might still build up to be significant.   This remains an
open question.

     The simulations we have done so far have all assumed {\it zero}
large-scale field, primarily as a result of our effort to minimize the
number of arbitrarily chosen free parameters.   Interestingly, we have
found that even when the magnetic field in the accreting matter has no
net flux at all, and therefore no externally-imposed connections to
infinity, under the right circumstances it can spontaneously {\it create}
such connections within a limited volume.

      To be specific, when the accretion flow contains closed dipolar field
loops large enough that the outer ends are accreted long ($\gg 10^3 GM/c^3$) after
the inner ones, a jet is automatically created from the flux provided by the
inner half of the field loop (\cite[McKinney \& Gammie 2004]{mg04},
\cite[Hawley \& Krolik 2006]{hk06}).  When flux is brought toward the black
hole along with the accretion flow, as soon as field lines begin to thread
the event horizon, matter drains off them, and the field lines, freed of the
matter's inertia, float upward.   Because a centrifugal barrier prevents any
matter with non-zero angular momentum from penetrating into a cone surrounding
the rotation axis, there is little inertia above these field lines.  The
field lines then rapidly expand upward.  This process of filling the region
around the rotation axis with magnetic field ceases only when there is
enough field intensity that the magnetic pressure distribution reaches equilibrium.
If the central black hole spins,
the portions of the field lines within the ergosphere are forced to rotate
along with the black hole, imposing a twist on the field lines.  The result
is Poynting flux travelling outward through the evacuated cone around the
rotation axis.

      When the black hole rotates rapidly, the electromagnetic luminosity
can be quite sizable.  Table~\ref{tab:bzpower} presents that luminosity,
normalized by the rest-mass accretion rate, as a function of black hole
spin.  The radiative efficiency predicted by the Novikov-Thorne model
is also given in that table in order to provide a standard of comparison.
As can be seen, when the black hole rotates rapidly, the jet efficiency becomes
comparable to the putative radiative efficiency.  Thus, consistent with
what many have long speculated, black hole rotation does seem to enhance
jet luminosity.

\begin{table}
  \begin{center}
  \begin{tabular}{ccc}
      \underbar{$a/M$}  & \underbar{$\eta_{EM}$}   &   \underbar{$\eta_{NT}$} \\[3pt]
      -0.9   & 0.023 & 0.039\\
      0.0    & 0.0003 & 0.057\\
      0.5    & 0.0063 & 0.081\\
      0.9    & 0.046 & 0.16\\
      0.93   & 0.038 & 0.17\\
      0.95   & 0.072 & 0.18\\
      0.99   & 0.21  & 0.26\\
  \end{tabular}
  \caption{Jet Poynting flux efficiency in rest-mass units $\eta_{EM}$, as a
  function of black hole spin parameter $a/M$ (\cite[Hawley \& Krolik 2006]{hk06}).
  These numbers can be compared
  with the radiative efficiency predicted by the Novikov-Thorne model, i.e.,
  the specific binding energy of a particle in the innermost stable circular
  orbit.
  \label{tab:bzpower}}
  \label{tab:kd}
  \end{center}
\end{table}

     On the other hand, black hole spin may not be the only relevant parameter.
For example, large dipolar loops are not the only imaginable field structure
for an accretion flow.  One could just as easily imagine narrower dipolar loops,
or quadrupolar loops (loops that don't cross the equatorial plane) or toroidal
loops.  These other geometries are in general less favorable to jet support
than the initial form explored, the large dipolar loops (Beckwith, Hawley
\& Krolik, in preparation).  Real systems are likely to exhibit some mixture
of these sorts of field topologies, and that mixture could easily vary from
one object to another, or from one time to another in a single object.  Some
of the observed variability in jets may conceivably reflect varying field
structures in the matter fed to the central black hole.

\section{Conclusions}\label{sec:concl}

    Thanks to the fundamental discovery that stresses in accretion disks come
from correlated MHD turbulence, driven by the magneto-rotational instability,
we can now begin to speak with confidence about a number of aspects of their
operation.  With the aid of ever-more-detailed and realistic numerical simulations,
we have taken the first steps toward connecting their internal dynamics with
observable properties.

    In this talk, advances in this direction in three areas have been reported:

\begin{itemize}

\item We now see that angular momentum transport is accomplished quasi-coherently,
by magnetic stress.  Because this mechanism is not a kinetic process
like viscosity, it is {\it not} intrinsically dissipative, although the associated
MHD turbulence does eventually dissipate.

    Moreover, far from ceasing at the innermost stable circular orbit, as has
been generally assumed for more than thirty years, magnetic stress continues
through the marginally stable region and deep into the plunging region.   When
the black hole spins, the stress can be continuous all the way to the event
horizon.  At the very least, the ability of electromagnetic stresses to carry
angular momentum away from the black hole and into the accretion flow means that
the spin-up rate of black holes can be rather less than would have been estimated
on the basis of accreting matter with the specific angular momentum of the
last stable orbit.   It is also possible, although quantitative determination
of this effect remains a job for the future, that these extended stresses lead
to extended dissipation as well, augmenting the radiative efficiency of black
hole accretion beyond the traditional values.

\item Detailed study of the vertical structure of disks subject to the MRI
has shown that their upper layers are supported primarily by magnetic pressure
gradients.  In addition, these upper layers can be far from the smooth time-steady
plane-parallel condition in which they are commonly imagined.  As a result,
the density at the photosphere is likely to be rather smaller than previously
estimated, and the locally-emitted spectrum may have significant departures
from black-body form.

    Ongoing work promises to clear up the long-standing mystery of whether
radiation-dominated disk regions are thermally unstable, and if they are,
what happens in the nonlinear stage of this instability.

\item As had been initially pointed out in the mid-1970s, when large-scale magnetic fields
pass close by rotating black holes, it is possible for very energetic relativistic
jets to be driven, deriving their energy from the rotational kinetic energy of
the black hole itself.   We can now begin to compute the detailed structure
of these jets, as functions of both space and time.   In addition, we now
see that it is possible to create large-scale magnetic field threading the ergosphere
of the black hole and stretching out to infinity from much smaller-scale field
embedded in the accretion flow---but not all small-scale field structures are
capable of doing this.

\end{itemize}

\begin{acknowledgments}

    I am indebted to my many collaborators on the work reported here.  In
alphabetical order, they are: Kris Beckwith, Omer Blaes, Shane Davis, Jean-Pierre
De~Villiers, John Hawley, Shigenobu Hirose, Ivan Hubeny, Yawei Hui, Jeremy Schnittman,
and Jim Stone.   This work was partially supported by NSF Grants AST-0313031 and
AST-0507455 and NASA ATP Grant NAG5-13228.

\end{acknowledgments}


\begin{thebibliography}{}

 \bibitem[Balbus \& Hawley (1991)]{bh91a}
     \textsc{Balbus, S.A. \& Hawley, J.F.} 1991
     {A powerful local shear instability in weakly magnetized disks. I - Linear analysis.}
     \textit{Ap. J.} \textbf{376}, 214--223

 \bibitem[Blaes et~al. (2006)]{bdhks06}
     \textsc{Blaes, O.M., Davis, S.W., Hirose, S., Krolik, J.H. \& Stone, J.M.} 2006
     {Magnetic Pressure Support and Accretion Disk Spectra}
     \textit{Ap. J.} \textbf{645}, 1402--1407

 \bibitem[Blaes et~al. (2007)]{bhk07}
     \textsc{Blaes, O.M., Hirose, S. \& Krolik, J.H.} 2007
     {Surface Structure in an Accretion Disk Annulus with Comparable Radiation and
      Gas Pressure}
     \textit{Ap. J.} \textbf{664}, 1057--1071

 \bibitem[Blandford \& Payne (1983)]{bp83}
     \textsc{Blandford, R.D. \& Payne, D.G.} 1983
     {Hydromagnetic flows from accretion discs and the production of radio jets}
     \textit{M.N.R.A.S.} \textbf{199}, 883--903

 \bibitem[Blandford \& Znajek (1977)]{bz77}
     \textsc{Blandford, R.D. \& Znajek, R.L.} 1977
     {Electromagnetic extraction of energy from Kerr black holes}
     \textit{M.N.R.A.S.} \textbf{179}, 433--456

 \bibitem[De~Villiers \& Hawley (2003)]{deVH03}
     \textsc{De~Villiers, J.-P. \& Hawley, J.F.} 2003
     {A Numerical Method for General Relativistic Magnetohydrodynamics}
     \textit{Ap. J.} \textbf{589}, 458--480

 \bibitem[Gammie (1999)]{g99}
     \textsc{Gammie, C.F.} 1999
     {Efficiency of Magnetized Thin Accretion Disks in the Kerr Metric}
     \textit{Ap. J. Letts.} \textbf{522}, L57--60

 \bibitem[Gammie et~al. (2003)]{gmt03}
     \textsc{Gammie, C.F., McKinney, J.C. \& T\'oth, G.} 2003
     {HARM: A Numerical Scheme for General Relativistic Magnetohydrodynamics}
     \textit{Ap. J.} \textbf{589}, 444--457

 \bibitem[Hawley \& Balbus (1991)]{bh91b}
     \textsc{Hawley, J.F. \& Balbus, S.A.} 1991
     {A powerful local shear instability in weakly magnetized disks.
      II - Nonlinear evolution.}
     \textit{Ap. J.} \textbf{376}, 223--233

 \bibitem[Hawley \& Krolik (2006)]{hk06}
      \textsc{Hawley, J.F. \& Krolik, J.H.} 2006
      {Magnetically Driven Jets in the Kerr Metric}
      \textit{Ap. J.} \textbf{641}, 103--116

 \bibitem[Hirose et~al. (2004)]{hkdh04}
     \textsc{Hirose, S., Krolik, J.H., De~Villiers, J.-P. \& Hawley, J.F.} 2004
     {Magnetically Driven Accretion Flows in the Kerr Metric. II. Structure of
     the Magnetic Field}
     \textit{Ap. J.} \textbf{606}, 1083--1097

\bibitem[Hirose et~al. (2006)]{hks06}
     \textsc{Hirose, S., Krolik, J.H. \& Stone, J.M.} 2006
     {Vertical Structure of Gas Pressure-dominated Accretion Disks with Local
     Dissipation of Turbulence and Radiative Transport}
     \textit{Ap. J.} \textbf{640}, 901--917

 \bibitem[Krolik (1999)]{k99}
     \textsc{Krolik, J.H.} 1999
     {Magnetized Accretion inside the Marginally Stable Orbit around a Black Hole}
     \textit{Ap. J. Letts.} \textbf{515}, L73--76

\bibitem[Krolik et~al. (2005)]{khh05}
     \textsc{Krolik, J.H., Hawley, J.F. \& Hirose, S.} 2005
     {Magnetically Driven Accretion Flows in the Kerr Metric. IV. Dynamical Properties
     of the Inner Disk}
     \textit{Ap. J.} \textbf{622}, 1008--1023

 \bibitem[Krolik et~al. (2007)]{khb07}
     \textsc{Krolik, J.H., Hirose, S. \& Blaes, O.M.} 2007
     {Thermodynamics of an Accretion Disk Annulus with Comparable Radiation and
     Gas Pressure}
     \textit{Ap. J.} \textbf{664}, 1045--1056

 \bibitem[McKinney \& Gammie (2004)]{mg04}
     \textsc{McKinney, J.C. \& Gammie, C.F.} 2004
     {A Measurement of the Electromagnetic Luminosity of a Kerr Black Hole}
     \textit{Ap. J.} \textbf{611}, 977--995

 \bibitem[Miller \& Stone (2000)]{ms00}
     \textsc{Miller, K.A. \& Stone, J.M.} 2000
     {The Formation and Structure of a Strongly Magnetized Corona above a
     Weakly Magnetized Accretion Disk}
      \textit{Ap. J.} \textbf{534}, 398--419

 \bibitem[Page \& Thorne (1974)]{pt74}
     \textsc{Page, D.N. \& Thorne, K.S.} 1974
      {Disk-Accretion onto a Black Hole. Time-Averaged Structure of Accretion Disk}
      \textit{Ap. J.} \textbf{191}, 499--506

 \bibitem[Phinney (1984)]{ph83}
     \textsc{Phinney, E.S.} 1983
     {unpublished Cambridge University Ph.D. thesis}

 \bibitem[Shakura \& Sunyaev (1973)]{ss73}
     \textsc{Shakura, N.I. \& Sunyaev, R.A.} 1973
     {Black holes in binary systems. Observational appearance.}
     \textit{Astron. Astrophys.} \textbf{24}, 337--355.

 \bibitem[Shakura \& Sunyaev (1976)]{ss76}
     \textsc{Shakura, N.I. \& Sunyaev, R.A.} 1976
     {A theory of the instability of disk accretion on to black holes and the
     variability of binary X-ray sources, galactic nuclei and quasars}
     \textit{M.N.R.A.S.} \textbf{175}, 613--632

\end{thebibliography}
\end{document}